\documentclass[12pt]{iopart}

\usepackage{graphicx}

\begin{document}

\title[Exactness of CVM for a protein folding model]{Exactness of the
cluster variation method and factorization of the equilibrium
probability for the Wako--Sait\^o--Mu\~noz--Eaton model of protein
folding}

\author{Alessandro Pelizzola\dag
\footnote[3]{alessandro.pelizzola@polito.it}
}

\address{\dag\ Dipartimento di Fisica, Politecnico di Torino, and
INFN, Sezione di Torino, c. Duca degli Abruzzi 24, 10129 Torino, Italy}

\begin{abstract}
I study the properties of the equilibrium probability distribution of
a protein folding model originally introduced by Wako and Sait\^o, and
later reconsidered by Mu\~noz and Eaton. The model is a
one--dimensional model with binary variables and many--body,
long--range interactions, which has been solved exactly through a
mapping to a two--dimensional model of binary variables with local
constraints. Here I show that the equilibrium probability of this
two--dimensional model factors into the product of local cluster
probabilities, each raised to a suitable exponent. The clusters
involved are single sites, nearest--neighbour pairs and square
plaquettes, and the exponents are the coefficients of the entropy
expansion of the cluster variation method. As a consequence, the
cluster variation method is exact for this model. 
\end{abstract}





\section{Introduction}

The cluster variation method (CVM) is an approximate method of
equilibrium statistical physics introduced by Kikuchi \cite{Kik51} as
a generalization of the Bethe--Peierls \cite{Bet35,Pei36} and
Kramers--Wannier \cite{KraWan1,KraWan2} approximations. In its modern
formulation \cite{An} it is based on the minimization of an
approximate variational free energy which is derived from the exact
one by a truncation of the cumulant expansion of the entropy. This
free energy depends on the probability distributions of local
clusters. An account of the CVM and its applications up to the
beginning of the '90s can be found in \cite{PTPS}. A review article is
also in preparation by the present author \cite{MyTopicalReview}.

Recently the relationship between the Bethe--Peierls approximation,
that is the lowest order CVM approximation, and the belief propagation
method \cite{Pearl}, widely used for inference and optimization
problems defined in terms of probabilistic graphical models, has been
discovered \cite{Yed01,KabSaa}, and this has led to the development of
the so--called generalized belief propagation (GBP) \cite{Yed01}, an
iterative message--passing algorithm whose fixed points are stationary
points of the CVM free energy.

There are only a few cases in which the CVM, and hence the GBP
algorithm, gives exact results. In most cases the exactness is due to
the tree--like topological structure of the underlying lattice or
graph. Leaving apart tree--like structures, the only case for which I
am aware of the exactness of the CVM is that of disorder varieties
\cite{Disorder1,Disorder2}, which occur in two--dimensional Ising
models with short--range competitive interactions on ordinary
translation--invariant lattices, where the correlations have a
particularly simple form which resembles that of one--dimensional
models. 

In all the cases in which the CVM is exact, independent of the reason,
the equilibrium probability distribution factors into a product of
local cluster probabilities, each raised to a suitable integer
exponent.

Given the limited number of cases in which the CVM gives an exact
solution, it is particularly relevant to show that for a particular
protein folding model the CVM is exact and the equilibrium
probability factors. This is the purpose of the present paper. 

The model I am going to study has been introduced and exactly solved
for the first time in 1978 by Wako and Sait\^o
\cite{WakoSaito1,WakoSaito2}, who showed its potential application to
the protein folding problem. Much later, Mu\~noz and Eaton
\cite{MunozEaton1,MunozEaton2,MunozEaton3} reconsidered this model, in
a slightly different formulation, solved it within the so-called
single (double, triple) sequence approximation, and made comparisons
with their own experimental data, after which the model became rather
popular (see for instance \cite{Popular}). Flammini, Banavar and
Maritan \cite{FlaBanMar} derived a mean--field theory, made Monte
Carlo simulations and solved exactly the $\alpha$--helix case and
later Bruscolini and I \cite{BruPel1,BruPel2} solved exactly the
general case, using ideas that are essentially the same as those by
Wako and Sait\^o \cite{WakoSaito1,WakoSaito2}. The same model and
techniques have recently been used for strained epitaxy on a modulated
substrate \cite{TokarDreysse}. 

In all these papers no reference was ever made to the original works
by Wako and Sait\^o. As far as I know, the first paper citing both
Wako--Sait\^o and Mu\~noz--Eaton is due to Itoh and Sasai
\cite{ItohSasai}, which refer to the model as the
Wako--Sait\^o--Mu\~noz--Eaton (WSME) model, as I'll do in the
following. 

The WSME model is a one--dimensional effective model, where a given
protein is regarded as a sequence of monomers (residues) connected by
peptidic bonds, and a binary variable is associated to each peptidic
bond, with the value 1 representing the folded (or native)
configuration and the value 0 representing the unfolded
configuration. An entropic cost is associated to the folded
configuration. Only native interactions are considered, like in G\^o
\cite{Go} models, that is two residues are allowed to interact only if
they are in contact in the native state. Moreover, and this is the
main characteristic of the present model, two residues can interact
only if all the peptidic bonds between them in the sequence are in the
native state, which gives rise to the many--body, long--range
interactions.

The exact solution of the equilibrium thermodynamics of the model can
be obtained through a mapping to a two--dimensional model
\cite{BruPel1}, defined on a triangular--shaped portion of the square
lattice, with local constraints as the only interactions. The
dimension of the state space of a row of this model is at most equal
to the length of the protein and this makes the transfer matrix
approach feasible.

In the present paper I show that the equilibrium probability
distribution of this model in its two--dimensional version can be
written in a factorized form which is typical of generalized
mean--field theories. It is a product of local cluster probabilities,
marginals of the global probability. In this product only
single--site, nearest--neighbour (NN) and square plaquette clusters appear,
and the corresponding probabilities are raised to exponents which are
the coefficients of the entropy expansion which appears in the
formulation of the CVM \cite{An}. As a
consequence, the CVM turns out to be exact (which was already noticed
empirically in \cite{BruPel1}), providing an exact variational free
energy depending on a number of variables which scales only
quadratically with the length of the protein.

The plan of the paper is as follows: in \Sref{Model} the model is
described in detail, its mapping to a two--dimensional model is
discussed in \Sref{Mapping}, then the factorization of the probability
distribution is shown in \Sref{Factor}. The CVM is described in
\Sref{CVM}, where its exactness for the present model is shown and the
consequences of this property are discussed. Finally, conclusions are
drawn in \Sref{Conclusions}.

\section{The model}
\label{Model}

The WSME model, if considered as a model for protein folding, is a
simplified effective model in the sense that one considers a very
restricted state space, and imagines to integrate with respect to all
the degrees of freedom except the binary peptidic bond variables,
obtaining a Hamiltonian which is actually an effective free energy. 

In order to describe the model, consider a protein of length $N+1$
residues, and associate a binary variable $m_i$, $i = 1, \ldots N$ to
each peptidic bond between residues $i$ and $i+1$. $m_i = 1$ denotes a
peptidic bond in a native configuration, $m_i = 0$ a peptidic bond in
an unfolded one. Since there are actually many more unfolded
configurations than native ones, an entropic cost $\Delta s_i < 0$ is
associated to each residue.

Two residues are allowed to interact only if they are in contact in
the native state. A detailed definition of contact between residues is
not needed here and can be found in
\cite{MunozEaton1,MunozEaton2,MunozEaton3}. Here I simply 
assume that a contact matrix $\Delta$ is given, with elements
$\Delta_{ij} = 1$ if residues $i$ and $j+1$ (or, equivalently,
peptidic bonds $i$ and $j$) are in contact in the native state, and 0
otherwise. 

As a further condition, residues $i$ and $j+1$ interact (with an
energy $\epsilon_{i,j}$) only if all 
the peptidic bonds between them are in the native state, that is only
if $\displaystyle{\prod_{k=i}^j} m_k = 1$. We can therefore write the
Hamiltonian (effective free energy)
\begin{equation}
H(\bi{m}) = \sum_{i=1}^{N-1} \sum_{j=i+1}^N \epsilon_{i,j} \Delta_{i,j}
\prod_{k=i}^j m_k - T \sum_{i=1}^N \Delta s_i m_i,
\end{equation}
where $T$ is the temperature.

The remainder of this paper does not focus on the applications to the
protein folding problem, but rather on the mathematical properties of
the statistical mechanics of the model itself, therefore from now on
we shall deal with a generic Hamiltonian of the form
\begin{equation}
H(\bi{m}) = \sum_{i=1}^{N} \sum_{j=i}^N h_{i,j} \prod_{k=i}^j m_k,
\label{Hgen}
\end{equation}
where the $h_{i,j}$'s can be temperature dependent. 

\section{Mapping to a 2--dimensional model}
\label{Mapping}

\begin{figure}
\begin{center}
\includegraphics*[scale=.9]{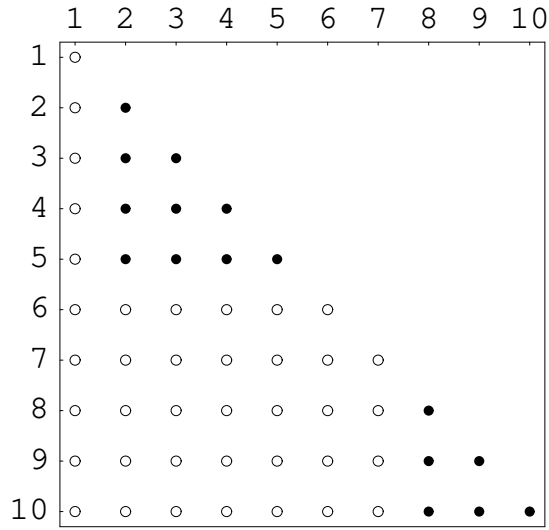}
\end{center}
\caption{\label{Fig2D}A typical configuration of the
  WSME model. An empty (resp.\ filled) circle at row $j$, column $i$
  represents the variable $m_{j,i}$ taking value 0 (resp.\
  1).}
\end{figure}

The above Hamiltonian suggests to introduce the new
binary variables \cite{BruPel1} $m_{j,i} =
\displaystyle{\prod_{k=i}^j} m_k$ (notice the order of the indices),
which take value 1 if $m_k = 1$, $i \le k \le j$, and 0
otherwise. These new variables can be associated to the
triangular--shaped portion of the square lattice defined by $1 \le i
\le j \le N$ and shown in \Fref{Fig2D}, and the original variables are
included in this set, since $m_{i,i} = m_i$.

The Hamiltonian can now be written as
\begin{equation}
H(\bi{m}) = \sum_{j=1}^{N} \sum_{i=1}^j h_{i,j} m_{j,i},
\label{H2D}
\end{equation}
but the new variables are not all independent. Since we have a model
with $2^N$ configurations written in terms of $N(N+1)/2$ binary
variables, constraints must exist between these variables. These
constraints can be written in different ways, and we chose
\cite{BruPel1} to write them as
\begin{equation}
m_{j,i} = m_{j-1,i} m_{j,i+1}, \qquad 1 \le i < j \le N.
\label{Constraints}
\end{equation}

One is therefore left with a model with a local (actually
non--interacting) Hamiltonian, and local constraints in place of the
interactions. Formally one could also include the constraints into the
Hamiltonian, by studying the limit $\lambda \to + \infty$ of
\begin{equation}
H(\bi{m}) = \sum_{j=1}^{N} \sum_{i=1}^j h_{i,j} m_{j,i} + \lambda
\sum_{j=2}^N \sum_{i=1}^{j-1} (m_{j,i} - m_{j-1,i} m_{j,i+1})^2. 
\label{H2DConstr}
\end{equation}

In the new representation the feasibility of the transfer matrix
approach becomes evident. Consider the variables in row $j$ in
\Fref{Fig2D}, that is $\bi{m}_j = (m_{j,1},m_{j,2}, \ldots
m_{j,j})$. Because of the constraints all configurations with $m_{j,i}
= 1$ and $m_{j,i+1} = 0$ are forbidden, leaving as the only allowed
states for $\bi{m}_j$ the $j+1$ states denoted by $\bi{e}_j^k$ and
defined by
\begin{equation}
m_{j,i} = \cases{
0, & $i \le k$ \\
1, & $i > k$ \\
}, \qquad 0 \le k \le j.
\label{RowStates}
\end{equation}
In \Fref{Fig2D} these states are readily depicted by placing $k$ 0's
on the left of row $j$ and $j-k$ 1's on the right.

A transfer matrix solution for a protein of length $N+1$ has then to
deal with the product of $N-1$ matrices, the largest of which has size
$(N+1) \times N$, the number of operations involved grows polynomially
in $N$, and the model can be easily solved even for long proteins.

\section{Factorization of the probability distribution}
\label{Factor}

The existence of a row--to--row transfer matrix is intimately
connected to the factorization of the model's probability distribution
over rows. This is well known in statistical mechanics
(see for instance \cite{Per77}). Readers interested in a
purely statistical formulation will regard this as an instance of the
junction tree theorem \cite{LauSpi,Jensen}. I report here the basic
steps of a proof for the present case.

First of all, write the Hamiltonian as a sum of row--to--row terms:
\begin{equation}
H(\bi{m}) = \sum_{j=1}^{N-1} H_{j,j+1}(\bi{m}_j, \bi{m}_{j+1}).
\end{equation}
If the Hamiltonian has the form reported in \Eref{H2DConstr}, then one
can write
\begin{equation}
\fl H_{j,j+1} = b_j \sum_{i=1}^j h_{i,j} m_{j,i} + b_{j+1}
\sum_{i=1}^{j+1} h_{i,j+1} m_{j+1,i} + \lambda
\sum_{i=1}^{j} (m_{j+1,i} - m_{j,i} m_{j+1,i+1})^2,
\end{equation}
where, in order to take into account the boundaries, $b_1 = b_N = 1$
and $b_j = 1/2$ for $1 < j < N$. For simplicity of notation, I drop
from now on the arguments of $H_{j,j+1}$. Of course, the specific form
of $H_{j,j+1}$ is irrelevant here, only the splitting into terms
involving only adjacent rows is important. 

Then we introduce the Boltzmann distribution 
\begin{equation}
p(\bi{m}) = \frac{1}{Z}\exp(-H) = \frac{1}{Z}\exp(-H_{1,2} \cdots
-H_{N-1,N}),   
\end{equation}
where we have absorbed $\beta = (k_B T)^{-1}$ in the definition of the
Hamiltonian and introduced the partition function
\begin{equation}
Z = \sum_{\bi{m}} \exp(-H) = \sum_{\bi{m}} \exp(-H_{1,2} \cdots
-H_{N-1,N}).
\end{equation}

Finally we define the marginal probability distributions for a row
\begin{equation}
p_j(\bi{m}_j) = \sum_{ \{ \bi{m}_l, l \ne j \} } p(\bi{m}), \qquad 1
\le j \le N, 
\end{equation}
and for a pair of adjacent rows
\begin{equation}
p_{j,j+1}(\bi{m}_j, \bi{m}_{j+1}) = \sum_{\{ \bi{m}_l, l \ne j, j+1 \}}
p(\bi{m}), \qquad 1 \le j \le N-1.
\end{equation}
The arguments of these probability will also be dropped in the
following.  

Introducing the partial partition functions for upper and
lower portions of the lattice
\begin{equation}
\eqalign{
Z_{U,j}(\bi{m}_j) &= \sum_{\bi{m}_{j-1}} \exp(-H_{j-1,j}) \cdots
\sum_{\bi{m}_1} \exp(-H_{1,2}), \\
Z_{L,j}(\bi{m}_j) &= \sum_{\bi{m}_{j+1}} \exp(-H_{j,j+1}) \cdots
\sum_{\bi{m}_N} \exp(-H_{N-1,N}), \\
}
\end{equation}
together with the boundary conditions
\begin{equation}
\eqalign{
Z_{U,1}(\bi{m}_1) &= 1, \\
Z_{L,N}(\bi{m}_N) &= 1, \\
}
\end{equation}
it is easy to show that
\begin{equation}
\eqalign{
p_j &= \frac{1}{Z} Z_{U,j}(\bi{m}_j) Z_{L,j}(\bi{m}_j), \\
p_{j,j+1} &= \frac{1}{Z} Z_{U,j}(\bi{m}_j) Z_{L,j+1}(\bi{m}_{j+1})
\exp(-H_{j,j+1}), \\ 
}
\end{equation}
and hence
\begin{equation}
\frac{p_{1,2} \cdots p_{N-1,N}}{p_2 \cdots p_{N-1}} = \frac{1}{Z}
\exp(-H) = p(\bi{m}).
\label{FactorTM}
\end{equation}
This shows that the global probability can be written as a product of
row--pair and row probabilities, raised to exponents 1 and -1
respectively, which is the first step towards our final result. As a
consequence, a CVM with all row--pairs as maximal clusters (see
\Sref{CVM}) would be exact for this model, but this is not interesting
here, since a much stronger property will be proved below. 

The above discussion is not specific to the present model and so far
the constraints have not yet been exploited. The next step is then to
make use of the constraints to write in a factor form similar to
\Eref{FactorTM} the row and row--pair probabilities.  

Consider first the row probability. Observe that, due to the
constraints \Eref{Constraints},
\begin{equation}
\eqalign{
m_{j,i} = 0 \qquad \Rightarrow \qquad m_{j,k} = 0 \quad \forall k < i,  \\
m_{j,i} = 1 \qquad \Rightarrow \qquad m_{j,k} = 1 \quad \forall k > i,  \\
}
\end{equation}
and the probability of the state $\bi{e}_j^k$ defined in \Eref{RowStates}
reduces to a NN pair probability:
\begin{equation}
p_j(\bi{e}_j^k) = p_j^{k,k+1}(0,1), \qquad 1 \le k < j
\end{equation}
where $p_j^{k,k+1}(m_{j,k},m_{j,k+1})$ is the NN (horizontal) pair
probability for the pair at row $j$ and columns $k, k+1$. Indeed,
$\bi{e}_j^k$ is the only state of row $j$ with $m_{j,k} = 0$ and
$m_{j,k+1} = 1$.

In addition, for the states $\bi{e}_j^0$ (all 1's) and $\bi{e}_j^j$
(all 0's) one can write
\begin{equation}
\eqalign{
p_j(\bi{e}_j^0) = p_j^{1,2}(1,1), \\
p_j(\bi{e}_j^j) = p_j^{j-1,j}(0,0) \\
}
\end{equation}
(actually the probabilities for these states could be reduced to site
probabilities, but this is not useful here).

Introducing the site probabilities $p_j^i(m_{j,i})$ and observing that
\begin{equation}
\eqalign{
p_j^{i,i+1}(0,0) = p_j^{i+1}(0), \\
p_j^{i,i+1}(1,1) = p_j^{i}(1), \\
}
\end{equation}
one can write the above results for the row probability in the
$k$--independent factorized form
\begin{equation}
p_j(\bi{e}_j^k) = \frac{p_j^{1,2}(m_{j,1},m_{j,2}) \cdots
p_j^{j-1,j}(m_{j,j-1},m_{j,j})}{p_j^{2}(m_{j,2}) \cdots
p_j^{j-1}(m_{j,j-1})}.
\label{FactorRow}
\end{equation}

A similar result can be obtained for the row--pair probability. In
this case one needs to define the square probability
$p_{j,j+1}^{i,i+1}(m_{j,i},m_{j,i+1};m_{j+1,i},m_{j+1,i+1})$, the
triangle probability
$p_{j,j+1}^{j,j+1}(m_{j,j};m_{j+1,j},m_{j+1,j+1})$ for the triangles
lying on the diagonal boundary and the NN (vertical) pair probability
$p_{j,j+1}^{i}(m_{j,i};m_{j+1,i})$.

There are two kinds of row--pair configurations:
$(\bi{e}_j^k,\bi{e}_{j+1}^k)$ which represents the cases in which
peptidic bonds $j$ and $j+1$ are in the same native stretch (e.g.\
rows 3 and 4 in \Fref{Fig2D}) or (for $k = j$) a new native stretch
starts at $j+1$ (e.g.\ rows 7 and 8 in \Fref{Fig2D}), and
$(\bi{e}_j^k,\bi{e}_{j+1}^{j+1})$ which represents the cases in which
a native stretch of length $j-k$ ends at $j$ (e.g.\ rows 5 and 6 in
\Fref{Fig2D}). For both kinds we have to show that a factorization
like \Eref{FactorRow} occurs.

The proof parallels the one for the row probability. Consider the
first kind of configurations and observe that, due to the constraints,
\begin{equation}
p_{j,j+1}(\bi{e}_j^k,\bi{e}_{j+1}^k) = p_{j,j+1}^{k,k+1}(0,1;0,1),
\qquad 1 \le k < j.
\label{RowPair1}
\end{equation}
For the boundary cases one has
\begin{equation}
\eqalign{
p_{j,j+1}(\bi{e}_j^0,\bi{e}_{j+1}^0) = p_{j,j+1}^{1,2}(1,1;1,1), \\
p_{j,j+1}(\bi{e}_j^j,\bi{e}_{j+1}^j) = p_{j,j+1}^{j,j+1}(0;0,1). \\
}
\label{RowPair2}
\end{equation}

Observing that
\begin{equation}
\eqalign{
p_{j,j+1}^{i,i+1}(0,0;0,0) = p_{j,j+1}^{i+1}(0;0) \\
p_{j,j+1}^{i,i+1}(1,1;1,1) = p_{j,j+1}^{i}(1;1) \\
}
\label{RowPair3}
\end{equation}
the row--pair probability can be written as 
\begin{equation}
p_{j,j+1}(\bi{e}_j^k,\bi{e}_{j+1}^k) = \frac{p_{j,j+1}^{1,2} \cdots
p_{j,j+1}^{j-1,j} p_{j,j+1}^{j,j+1}}{p_{j,j+1}^{2} \cdots
p_{j,j+1}^{j}} \qquad 0 \le k \le j,
\label{RowPair4}
\end{equation}
where the last factor in the numerator is a triangle probability. For
simplicity, the arguments of the probability have been dropped. 

Consider now configurations of the second kind. Equations
(\ref{RowPair1}), (\ref{RowPair2}) and (\ref{RowPair4}) are replaced by
\begin{equation}
p_{j,j+1}(\bi{e}_j^k,\bi{e}_{j+1}^{j+1}) = p_{j,j+1}^{k,k+1}(0,1;0,0),
\qquad 1 \le k < j,
\end{equation}
\begin{equation}
\eqalign{
p_{j,j+1}(\bi{e}_j^0,\bi{e}_{j+1}^{j+1}) = p_{j,j+1}^{1,2}(1,1;0,0), \\
p_{j,j+1}(\bi{e}_j^j,\bi{e}_{j+1}^{j+1}) = p_{j,j+1}^{j,j+1}(0;0,0) \\
}
\end{equation}
and
\begin{equation}
p_{j,j+1}(\bi{e}_j^k,\bi{e}_{j+1}^{j+1}) = \frac{p_{j,j+1}^{1,2} \cdots
p_{j,j+1}^{j-1,j} p_{j,j+1}^{j,j+1}}{p_{j,j+1}^{2} \cdots
p_{j,j+1}^{j}} \qquad 0 \le k \le j,
\end{equation}
respectively. 

The factorization property
\begin{equation}
p_{j,j+1}(\bi{m}_j,\bi{m}_{j+1}) = \frac{p_{j,j+1}^{1,2} \cdots
p_{j,j+1}^{j-1,j} p_{j,j+1}^{j,j+1}}{p_{j,j+1}^{2} \cdots
p_{j,j+1}^{j}} \qquad 0 \le k \le j
\label{FactorRowPair}
\end{equation}
is then proved for any valid row--pair configuration. 

In order to obtain the final result of this section, plug Equations
(\ref{FactorRow}) and (\ref{FactorRowPair}) into \Eref{FactorTM},
obtaining 
\begin{equation}
p(\bi{m}) = \frac{ \left( \displaystyle{\prod_{j=1}^{N-1} \prod_{i=1}^j}
  p_{j,j+1}^{i,i+1} \right) 
\left( \displaystyle{\prod_{j=3}^{N-1} \prod_{i=2}^{j-1}} p_j^i \right) }
{ \left( \displaystyle{\prod_{j=2}^{N-1} \prod_{i=2}^j} p_{j,j+1}^i \right) 
\left( \displaystyle{\prod_{j=2}^{N-1} \prod_{i=1}^{j-1}} p_j^{i,i+1} 
\right) },
\label{FactorFull}
\end{equation}
that is the global probability is reduced to a product of square,
triangle and site probabilities divided by a product of pair
probabilities. 

\section{Exactness of the cluster variation method}
\label{CVM}

In this section, after a brief introduction to the CVM, it is shown
that the factorization \Eref{FactorFull} implies the exactness of the
CVM, and the consequences of this property are discussed.

The CVM in its modern formulation is derived from the variational
principle of statistical mechanics, which states that, given a model
with variables $\bi{m}$ and Hamiltonian $H(\bi{m})$, its equilibrium
(Boltzmann) probability distribution $p_{\rm eq}(\bi{m}) =
\exp(-H(\bi{m}))/Z$ is the distribution which minimizes the
variational free energy
\begin{equation}
F = U - S = \sum_\bi{m} p(\bi{m}) H(\bi{m}) + \sum_\bi{m} p(\bi{m})
\ln p(\bi{m}).
\label{VarF}
\end{equation}
The physical values of the free energy $F$, the energy $U$ and the
entropy $S$ can then be calculated at the minimum. 

If the variables are associated to the nodes of a graph, or to the
sites of a lattice, one can define clusters of nodes (e.g.\ single
sites, pairs, triangles, square plaquettes, \ldots). For each
cluster $\alpha$ define also the corresponding set of variables
$\bi{m}_\alpha$, probability distribution $p_\alpha(\bi{m}_\alpha)$
and entropy
\begin{equation}
S_\alpha = - \sum_{\bi{m}_\alpha} p_\alpha(\bi{m}_\alpha) \ln
p_\alpha(\bi{m}_\alpha).
\end{equation}

The entropy cumulants are defined by
\begin{equation}
S_\alpha = \sum_{\beta \subseteq \alpha} \tilde S_\beta,
\end{equation}
which can be solved with respect to the cumulants by means of a
M\"obius inversion, which yields
\begin{equation}
\tilde S_\beta = \sum_{\alpha \subseteq \beta} (-1)^{n_\alpha -
  n_\beta} S_\alpha,
\end{equation}
where $n_\alpha$ denotes the number of nodes in cluster $\alpha$. 

The full entropy 
\begin{equation}
S = \sum_\beta \tilde S_\beta
\end{equation}
can then be approximated by selecting a set $R$ of clusters, made of
certain maximal clusters and all their subclusters, and truncating the
cumulant expansion by retaining only terms which correspond to
clusters in $R$. One obtains
\begin{equation}
S \simeq \sum_{\beta \in R}  \tilde S_\beta = \sum_{\alpha \in R}
a_\alpha S_\alpha,
\end{equation}
where the coefficients $a_\alpha$, sometimes called M\"obius numbers,
satisfy \cite{An}
\begin{equation}
\sum_{\alpha \in R, \alpha \supseteq \beta} a_\alpha = 1 \qquad
\forall \beta \in R.
\label{MobiusNum}
\end{equation}
The above condition, practically useful for determining the
$a_\alpha$'s, means that every subcluster must be counted exactly once
in the entropy expansion. 

Now assume that the Hamiltonian is made of local terms only, and $R$
has been chosen such that one can write
\begin{equation}
H(\bi{m}) = \sum_{\alpha \in R} H_\alpha(\bi{m}_\alpha).
\label{HCVM}
\end{equation}
In such a case the variational free energy can be written, with
the above approximation on the entropy and no approximation on the
energy, as
\begin{equation}
F_{CVM} = \sum_{\alpha \in R} \sum_{\bi{m}_\alpha}
p_\alpha(\bi{m}_\alpha) H_\alpha(\bi{m}_\alpha) + \sum_{\alpha \in R}
a_\alpha \sum_{\bi{m}_\alpha} p(\bi{m}_\alpha) \ln p(\bi{m}_\alpha),
\label{VarFCVM}
\end{equation}
where the minimization must be performed with respect to the
$p_\alpha$'s with the constraints
\begin{equation}
\eqalign{
\sum_{\bi{m}_\alpha} p_\alpha(\bi{m}_\alpha) = 1, \qquad \forall \alpha
\in R, \\
\sum_{\bi{m}_{\alpha \setminus \beta}} p_\alpha(\bi{m}_\alpha) =
p_\beta(\bi{m}_\beta), \qquad \forall \beta \subset \alpha \in R. \\
}
\end{equation}

It is interesting to observe that a suitable factorization of the equilibrium
probability of the model implies that the variational free energy
\Eref{VarF} reduces to the CVM variational free energy
\Eref{VarFCVM} with no approximation. More precisely, assume that the
equilibrium probability factorizes in terms of its marginals according to
\begin{equation}
p_{\rm eq}(\bi{m}) = \prod_{\alpha \in R} \left[ p_\alpha(\bi{m}_\alpha)
  \right]^{a_\alpha}.
\label{FactorCVM}
\end{equation}
Then one can restrict the variational principle to distributions
$p(\bi{m})$ with the same property. The entropy of such a distribution
is 
\begin{equation}
\eqalign{
S & = - \sum_{\bi{m}} p(\bi{m}) \ln p(\bi{m}) = \\
& = - \sum_\alpha a_\alpha
\sum_{\bi{m}} p(\bi{m}) \ln p_\alpha(\bi{m}_\alpha) = \\
& = - \sum_\alpha
a_\alpha \sum_{\bi{m}_\alpha} p_\alpha(\bi{m}_\alpha) \ln
p_\alpha(\bi{m}_\alpha),
}
\end{equation}
and the CVM free energy is therefore obtained with no approximation
(recall that no approximation was made on the energy term).

The WSME model in its two--dimensional representation falls precisely
in this case. If one chooses as
maximal clusters all the square plaquettes and the triangles lying on
the diagonal boundary the Hamiltonian can obviously be written as in
\Eref{HCVM} and the M\"obius numbers for the entropy expansion,
obtained by \Eref{MobiusNum}, are: 
\begin{description}
\item[$\bullet$] $a_\alpha = 1$ for the maximal clusters;
\item[$\bullet$] $a_\alpha = 0$ for the triangles not lying on the
diagonal boundary and the next--nearest--neighbour pairs, since they
are contained in exactly 1 maximal cluster;
\item[$\bullet$] $a_\alpha = -1$ for all the NN pairs contained in 2
maximal clusters, that is all NN pairs except the boundary ones;
\item[$\bullet$] $a_\alpha = 0$ for the boundary NN pairs;
\item[$\bullet$] $a_\alpha = 1$ for the sites contained in 4 maximal
clusters and 4 NN pairs, that is all sites except the boundary ones;
\item[$\bullet$] $a_\alpha = 0$ for the boundary sites.
\end{description}

With the above M\"obius numbers it is immediate to check that the
factorization \Eref{FactorCVM} is exactly the one that we have
obtained in \Eref{FactorFull} for the WSME model, and hence the CVM is
exact for this model. The corresponding variational free energy,
dropping all the arguments in the entropy term, reads
\begin{equation}
\eqalign{ 
F = & \sum_{j=1}^N \sum_{i=1}^j h_{i,j} \sum_{m_{j,i}}
m_{j,i} p_j^i(m_{j,i}) + \sum_{j=1}^{N-1} \sum_{i=1}^j
p_{j,j+1}^{i,i+1} \ln p_{j,j+1}^{i,i+1} + \\ 
& - \sum_{j=2}^{N-1}
\sum_{i=2}^j p_{j,j+1}^{i} \ln p_{j,j+1}^{i} - \sum_{j=2}^{N-1}
\sum_{i=1}^{j-1} p_{j}^{i,i+1} \ln p_{j}^{i,i+1} + \\
& + \sum_{j=3}^{N-1} \sum_{i=2}^{j-1} p_{j}^{i} \ln p_{j}^{i}.
}
\label{FpCVM}
\end{equation}
The local constraints \Eref{Constraints} can be either included in the
Hamiltonian or imposed by hand on the probabilities. 

The numerical minimization of the above variational free energy can in
principle be performed by a provably--convergent, double--loop
algorithm like the one proposed by Heskes, Albers and Kappen
\cite{HAK}, while the GBP fails, probably due to the constraints,
except in very simple cases. This is not an important point however,
since the strength of this result is not that it improves on the
transfer matrix method, which is already very efficient in solving for
the equilibrium. On this side the only advantage of this approach is
probably that in this scheme it is easier to calculate correlation
functions, since the probabilities are directly accessible. 

The real strength of this result, apart from having found a new model
which is exactly solvable by the CVM, is that here the CVM can serve
as a basis to build powerful approximation for the dynamics of the
model, which is extremely relevant for the protein folding
problem. For this purpose, it might also be useful to observe that the
variational free energy can be written explicitly in terms of the
local expectations
\begin{equation}
x_{j,i} = \langle m_{j,i} \rangle = \sum_{m_{j,i} = 0, 1} m_{j,i}
p_j^i(m_{j,i}) = p_j^i(1).
\label{LocalExp}
\end{equation}
To this end observe that in the variational free energy \Eref{FpCVM}
the energy term is already a linear function of these expectations,
while the probabilities appearing in the entropy term can be written
as functions of these expectations as independent variables. For
instance, using \Eref{LocalExp} and normalization one obtains
\begin{equation}
\eqalign{
p_j^i(1) = x_{j,i} \\
p_j^i(0) = 1 - x_{j,i}. \\
}
\end{equation}
For the NN pair one configuration is forbidden by the constraints, and
the remaining probabilities are determined by normalization and
marginalization to the site probabilities, obtaining
\begin{equation}
\eqalign{
p_j^{i,i+1}(0,0) = 1 - x_{j,i+1} \\
p_j^{i,i+1}(0,1) = x_{j,i+1} - x_{j,i} \\
p_j^{i,i+1}(1,1) = x_{j,i} \\
}
\end{equation}
for the horizontal pair and
\begin{equation}
\eqalign{
p_{j,j+1}^i(0,0) = 1 - x_{j,i} \\
p_{j,j+1}^i(1,0) = x_{j,i} - x_{j+1,i} \\
p_{j,j+1}^i(1,1) = x_{j+1,i} \\
}
\end{equation}
for the vertical pair. Similarly, for a triangle lying on the diagonal
boundary, only 4 configurations are allowed by the constraints, and
their probabilities are again determined by normalization and
marginalization to subclusters, with the result
\begin{equation}
\eqalign{
p_{j,j+1}^{i,i+1}(0;0,0) = 1 - x_{j,i} - x_{j+1,i+1} + x_{j+1,i} \\
p_{j,j+1}^{i,i+1}(0;0,1) = x_{j+1,i+1} - x_{j+1,i} \\
p_{j,j+1}^{i,i+1}(1;0,0) = x_{j,i} - x_{j+1,i} \\
p_{j,j+1}^{i,i+1}(1;1,1) = x_{j+1,i}. \\
}
\end{equation}
Finally, for a square plaquette we have 5 allowed configurations and
the probabilities
\begin{equation}
\eqalign{
p_{j,j+1}^{i,i+1}(0,0;0,0) = 1 - x_{j,i+1} \\
p_{j,j+1}^{i,i+1}(0,1;0,0) = x_{j,i+1} + x_{j+1,i} - x_{j,i} - x_{j+1,i+1} \\
p_{j,j+1}^{i,i+1}(0,1;0,1) = x_{j+1,i+1} - x_{j+1,i} \\
p_{j,j+1}^{i,i+1}(1,1;0,0) = x_{j,i} - x_{j+1,i} \\
p_{j,j+1}^{i,i+1}(1,1;1,1) = x_{j+1,i}. \\
}
\end{equation}
Substituting the above probabilities into the variational free energy
\Eref{FpCVM} yields an exact variational free energy as a function of
$N(N+1)/2$ independent local variables. 

\section{Conclusions}
\label{Conclusions}

I have shown that the equilibrium probability factors into a product
of local cluster probabilities, and hence the CVM is exact, for the
WSME model of protein folding. After the one--dimensional WSME model,
which has long--range, many--body interactions, has been mapped into a
two--dimensional model which has local constraints as the only
interactions, the proof goes through two steps. The first step
exploits the locality of the interactions, the second one the detailed
form of the constraints.

The result is especially relevant on the methodological side, since
leaving apart tree--like models, the CVM is exact, as far as I know,
only on disorder varieties of two--dimensional models. 

Moreover, some consequences can be expected also on the model side. As
far as equilibrium properties are concerned, we have almost no
improvement with respect to the transfer matrix method, which is
already very fast in this matter. The main advantage of the CVM
solution is that correlation functions are much easier to
calculate. The most important point is however that the CVM solution
for the equilibrium can be a starting point for good approximations
for the dynamics, which is of utmost importance in the context of the
protein folding problem. Two approaches can be followed
\cite{Tesi,Prep}. On one hand, one can develop a master equation
approach based on the approximation that the state of the system at
any time can be described as an equilibrium state of the WSME model,
with a time--dependent Hamiltonian (the so--called local equilibrium
approximation \cite{Kawasaki}). On the basis of the results reported
here, this means that the probability at any time is assumed to
factorize as the equilibrium one. Alternatively, one can build an
approximation for the dynamics with the path probability method (PPM,
the dynamical version of the CVM) \cite{PTPS}, with square plaquettes
and triangles as maximal clusters. Since the stationary state of the
PPM corresponds to the CVM solution for the equilibrium, one obtains
an approximation for the dynamics which is guaranteed to converge to
the exact equilibrium state. It can be verified \cite{Tesi,Prep} that
the two approaches are equivalent in the limit of vanishing time step
and that very good agreement is obtained with respect to the exact
solution for short chains.

\end{document}